# An Advanced Approach On Load Balancing in Grid Computing

**Dushyant Vaghela[1]**
Department of Computer Engineering
RK University,
Rajkot, India.

*Abstract: With the rapid development in wide area networks and low cost, powerful computational resources, grid computing has gained its popularity. With the advent of grid computing, space limitations of conventional distributed systems can be overcome and underutilized computing resources at different locations around the world can be put to distributed jobs. Workload and resource management is the main key grid services at the service level of grid infrastructures, out of which load balancing in the main concern for grid developers. It has been found that load is the major problem which server faces, especially when the number of users increases. A lot of research is being done in the area of load management. This paper presents the various mechanisms of load balancing in grid computing so that the readers will get an idea of which algorithm would be suitable in different situations.*

*Keywords: wide area network, distributed computing, load balancing.*

I. INTRODUCTION

A computational grid is a large scale, heterogeneous collection of autonomous systems, geographically distributed and interconnected by heterogeneous networks.

A grid can be defined as a large-scale geographically distributed hardware and software infra-structure composed of heterogeneous networked resources owned and shared by multiple administrative organizations which are coordinated to provide transparent, dependable, pervasive and consistent computing support to a wide range of applications. These applications can perform either distributed computing, high throughput computing, on-demand computing, data-intensive computing, collaborative computing or multimedia computing.[1][7]

Grid computing is a collection of computer resources from different geographical location to achieve a common goal. Computation grid uses network and combines computational resources from different geographical locations for distributed jobs.

Job sharing (computational burden) is one of the major difficult tasks in a computational grid environment. Grid resource manager provides the functionality for discovery and publishing of resources as well as scheduling, submission and monitoring of jobs. However, computing resources are geographically distributed under different ownerships each having their own access policy, cost and various constraints.

The authors in [1] have identified main characteristics the grid computing should have as follows:

Large scale: a grid must be able to deal with a number of resources ranging from just a few to millions. This raises the very serious problem of avoiding potential performance degradation as the grid size increases.

Geographical distribution: grid's resources may be located at distant places.

Heterogeneity: a grid hosts both software and hardware resources that can be very varied ranging from data, files, software components or programs to sensors, scientific instruments, display devices, personal digital organizers, computers, supercomputers and networks.

Resource sharing: resources in a grid belong to many different organizations that allow other organizations (i.e. users) to access them. Nonlocal resources can thus be used by applications, promoting efficiency and reducing costs.

Multiple administrations: each organization may establish different security and administrative policies under which their owned resources can be accessed and used. As a result, the already challenging network security problem is complicated even more with the need of taking into account all different policies.

> Resource coordination: resources in a grid must be coordinated in order to provide aggregated computing capabilities.

> Transparent access: a grid should be seen as a single virtual computer.

Dependable access: a grid must assure the delivery of services under established Quality of Service (QoS) requirements. The need for dependable service is fundamental since users require assurances that they will receive predictable, sustained and often high levels of performance.

Consistent access: a grid must be built with standard services, protocols and inter-faces thus hiding the heterogeneity of the resources while allowing its scalability. Without such standards, application development and pervasive use would not be possible.

*Pervasive access*: the grid must grant access to available resources by adapting to a dynamic environment in which resource failure is commonplace. This does not imply that resources are everywhere or universally available but that the grid must tailor its behavior as to extract the maximum performance from the available re-sources.

All the resource available in the grid environment are basically owned and managed by multiple organizations [2]. For the efficient operation of computational grid, various factors must be considered such as resources sharing, and load balancing. The primary challenges that should be taken into account for grid system are well identified in [2], and are as follow: ③ Administration & Security

- ③ Solution Development
- ③ Resource heterogeneity
- ③ Accounting infrastructure
- ③ Resource Management

For the deployment of various high-performance computing applications, the computational grid provides a promising platform. To achieve the high performance, an efficient scheduling of task onto the processors that minimizes the entire execution time is vital. It has been found that solving this problem is very hard and many attempts have been made.

A computational grid is a shared environment implemented via the deployment of a persistent, standards-based service infrastructure that supports the creation of, and resource sharing within, distributed communities. Resources can be computers, storage space, instruments, software applications, and data, all connected through the Internet and a middleware software layer that provides basic services for security, monitoring, resource management, and so forth. Resources owned by various administrative organizations are shared under locally defined policies that specify what is shared, who is allowed to access what, and under what conditions [48].

The real and specific problem that underlies the Grid concept is coordinated resource sharing and problem solving in dynamic, multi-institutional virtual organizations [44].

From the point of view of scheduling systems, a higher level abstraction for the Grid can be applied by ignoring some infrastructure components such as authentication, authorization, resource discovery and access control.

We consider the term Grid as *A type of parallel and distributed system that enables the sharing, selection, and aggregation of geographically distributed autonomous and heterogeneous resources dynamically at runtime depending on their availability, capability, performance, cost, and users' quality-of-service requirements*" [7]. We use the following terms as defined in [6]:

- A *task* is an atomic unit to be scheduled by the scheduler and assigned to a resource.
- The *properties* of a task are parameters like CPU/memory requirement, deadline, priority, etc.
- A *job* (or *metatask*, or *application*) is a set of atomic tasks that will be carried out on a set of resources. Jobs can have a recursive structure, meaning that jobs are composed of sub-jobs and/or tasks, and sub-jobs can themselves be decomposed further into atomic tasks. In this paper, the term *job*, *application* and *metatask* are interchangeable.
- A *resource* is something that is required to carry out an operation, for example: a processor for data processing, a data storage device, or a network link for data transporting.
- A *site* (or *node*) is an autonomous entity composed of one or multiple resources.
- A *task scheduling* is the mapping of tasks to a selected group of resources which may be distributed in multiple administrative domains.

## II. LOAD BALANCING

A load is the number of jobs in the waiting queue and can be light, moderate and heavy according to their work. Load balancing is a process of improving the performance of computational grid system in such a way that all the computing nodes involved in the grid are uniformly utilized as much as possible so that the throughput is improved, and the execution times are minimized.

Broadly the load balancing algorithms fall into two classes: static and dynamic. In static load balancing the decision information are made in advance. The prior knowledge of all the information related to the scheduling is known in static algorithm. Whereas in dynamic load balancing the scheduling decisions are made when there is need to schedule the job for further processing. A dynamic task scheduling can use either centralized or distributed control. In a centralized approach, all scheduling decision are made at one site and the failure in central site cause entire system down. In a distributed or noncooperative approach, each site makes its own scheduling decisions and the control is much scalable as well as more reliable [26].

This paper presents a review of different approaches proposed for the load balancing algorithms.

### III. VARIOUS LOAD BALANCING APPROACHES

For a dynamic load balancing algorithm, it is not acceptable to frequently change state information because of the high communication overheads. In [3] an estimated load information scheduling algorithm (ELISA) and Perfect Information Algorithm (PIA) is proposed. In PIA, when a job arrives, a processor computes the jobs finish time on all buddy processors using exact information about the current load of a buddy processor, its arrival rate and service rate. The source processor selects a buddy processor with the minimum finish time and immediately migrate a job on that buddy processor, if it can finish the job earlier than this processor. In the decentralized load balancing algorithm proposed in [5] for a Grid environment. Although this work attempts to include the communication latency between two nodes during the triggering process on their model, it did not consider the actual cost for a job transfer. In [6, 7], a sender processor collects status information about neighboring processors by communicating every load balancing moment. This can lead to frequent message transfers that results in large communication overhead which is undesirable. Preemptive and non-preemptive process migration techniques are proposed in [4]. In this process migration takes place efficiently by considering memory usage and load on processor Authors in [13], take the idea of assigning a priority to each computing node in the grid system based on their computing power. The algorithm proposed in [13], a migrating server node (MSN) returns light weighted node whenever required. It is done by checking the status of all the nodes which are under-loaded. When a node is overloaded, it calls the MSN which then finds a suitable node and then performs the load balancing. This approach is based on the principle that in a distributed environment at least one node should be lightly loaded. This approach considers the CPU queue length.

The existing load balancing algorithms usually find the under-loaded node by their status information exchange between the nodes. The algorithm works by defining a function called msn() which finds the available under-loaded nodes by looking into a queue where all the processors are scheduled in the decreasing order of their computing power. Hence, the probability that first node in the queue will be the under-loaded node is high as the first node is having the highest computing power and it may assume that it has finished the assigned work and can be idle that time, if not msn() will check the second node having second highest computing power and so on. Hence, it reduces the communication overhead as compare to other existing algorithms in which it is necessary to collect the status information of all the nodes to find out which node is under-loaded [13].

In [3] the authors have used the Genetic Algorithm to schedule the jobs in a Grid and found that the finishing times are improved. Authors could correctly apply and implement all the 8 components of GA in the context of Grid computing scheduling. The components are the Representation, Fitness function, Parent Selection mechanism, Reproduction, Cross over operator, Mutation operators, Survivor selection mechanism, and termination condition of ending the execution of GA.

Hongtu in his thesis [14] describes a distributed dynamic scheduling algorithm of composite tasks on a grid computing system. The algorithms include two steps: external scheduling and internal scheduling. The factors analyzed in the simulation experiments are the number of task submissions, the task submission interval, the machine failure rate, the network infrastructure, and SWQ size.

It has been found that the number of task submissions definitely affects network-computing performance. As more tasks are submitted, the network-computing performance drops because more tasks are competing for a fixed number of computing resources. Further the task submission intervals affects the network-computing performance. If a task submission interval is too short, it will cause local bidding errors; conversely, as the task submission interval increases, the number of remote bidding errors rises [14]. A decrease in the machine failure rate helped augment computing performance because it gave network computing more computing resources and minimized the impact that machine failure has on rescheduling and re-mapping. Author claims that if the network infrastructure does not affect other environmental parameters a 1-LAN structure is superior to a 100-LAN structure. Subtask Waiting Queue-SWQ size had impacts network computing performance, choosing the best SWQ size was a balance between the negative effects of having a size that was too long or too short. Authors found that the impact of number of submitted tasks, task submission intensity, reliability of computer system, and network infrastructure should be considered when submitting the tasks.

In [16], authors presented a new fuzzy logic-based scheme for dynamic load balancing in the grid computing services. To evaluate the node's workload, four parameters, Node's ready queue length, burst time, CPU utilization, and the available resources needed to accomplish the assigned tasks, have been considered, for the sake of more accuracy. Moreover, two fuzzylogic based models are used in different levels which are node-level and cluster- level. The simulation results show that the proposed scheme achieves really satisfactory and consistently load balancing than in other randomized approaches.

A mathematical model is presented in [17] to prove the load balance performance of the agent based resource scheduling algorithm. The authors have used Grid-P2P environment for the resource management. Utilizing this algorithm, the idle computational resources of DDG can be dynamically scheduled according to the real-time working load of each execution node.

A new long-term, application-level prediction model is proposed in [18]. This prediction model is derived from a combination of rigorous mathematical analysis and intensive simulation. Two important issues are handled in [18]. First, the prediction of job computation time in a non-dedicated network, by adopting a newly proposed long-term, application-level prediction model. This prediction model is derived from a combination of rigorous mathematical analysis and intensive simulation. The effects of machine utilization, computing power, and local job service and task allocation on the completion time of remote task are individually identified. Formulas to distinguish the impact of different factors are derived in the model analysis.

This model also addressed the match of QoS request from the application and the QoS provided by the diverse resources in Grid. The QoS information is embedded into the scheduling algorithm making a better match among different level of QoS request/supply. Consequently, the new scheduling algorithm improves the efficiency and the utilization of a Grid system.

Authors in [21] proposed a Load balancing algorithm for optimal scheduling. This algorithm schedules the task by minimum completion time and reschedules it by waiting time of each task to obtain load balance. The proposed scheduler completed a task by using highly utilized low cost resources with minimum computational time. The scheduling algorithm uses the priority queue of resources to achieve a higher throughput.

In [22] a distributed load balancing model is proposed which transforms any Grid topology into a forest structure and a two level strategy is developed to balance the load among resources of computational Grid. This strategy privileges local rather than global load balancing in order to achieve two main objectives:

I. The reduction of the average response time of tasks;

II. The reduction of communication cost induced by the task transferring.

This paper [22] takes into account the heterogeneity of the resources that is completely independent from any physical architecture Grid while load balancing.

The strategy is fully distributed, uses a task-level load balancing and privileges, as much as possible, a local load balancing to avoid the use of WAN communication. In order to validate the proposed load balancing strategy, authors developed two algorithms: intra-cluster and inter-clusters.

It is found that the strategy appreciably improves the average response time with a low communication cost.

We present in table 1 our observations.

TABLE 1

| Algorithm Type | Reference | Key concept / Algorithm used | Issues addressed | Remarks |
|---|---|---|---|---|
| Static | [3] | Genetic Algorithm | The new string representation has been used, communication costs has not been ignored. | Overall execution time is improved, reaching optimum solution. |
| Dynamic | [13] | Idea of assigning a priority to each computing node in the grid system based on their computing power. A migrating server node finction returns light weighted node whenever required. | Assignment of priority to each computing node. Communication overhead. | Algorithm reduces the communication overhead. Complexity of the proposed approach for load balancing is not experimentally proved. |
| Dynamic | [14] | Distributed Dynamic Scheduling of Composite Tasks. Tasks consists of subtasks represented by DAGs. External scheduling and mapping are performed on the task level, and internal scheduling and mapping are done on the subtask level. | Subtask Waiting Queue size, submitted task number, task submission interval, and network infrastructure. The percentage of tasks completed before deadline and average response times are used as indexes of network computing performance. | The task submission interval has an impact on network computing performance. Use of a near-future network computing load, during external scheduling and mapping, helps to eliminate the problems associated with local and remote bidding errors. How the nodes of a LAN can affect the scheduling and mapping time is not studied. |
| Dynamic | [16] | Fuzzy-logic-based scheme for dynamic load balancing in grid computing services | The performance is evaluated in terms of its ability to keep all nodes and clusters of the overall system in a balanced way. | Use of clustering for the nodes in the same cluster based on their states since enormous number of nodes may impair the proposed scheme |

| | | | | performance. Validation of the scalability of the proposed framework. |
|---|---|---|---|---|
| Dynamic | [17] | Agent based scheduling algorithm, utilizing the advantages of master-slave structure, P2P technology and agent knowledge. | -load balancing -robustness ( ability to regain consistence after node failures) | -good performance at load balance. -robustness is good |
| Dynamic | [18] | General adaptive heuristic scheduling | Security, Quality of Service (QOS), Central control within distributed administrative domains Prediction of job computation time in non dedicated network. | Improved performance (upto 12%) gain for a variety of applications. |
| Dynamic / fully distributed | [22] | load balancing model is proposed which transforms any Grid topology into a forest structure and a two level strategy is developed to balance the load among resources of computational Grid | The heterogeneity of the resource is taken into account. | Appreciable improvement of the average response time with a low communication cost. the effectiveness of algorithm is not tested in other simulators. |

## IV. CONCLUSION

The version of this template is V2. Most of the formatting instructions in this document have been compiled by Causal Productions from the IEEE LaTeX style files. Causal Productions offers both A4 templates and US Letter templates for LaTeX and Microsoft Word. The LaTeX templates depend on the official IEEEtran.cls and IEEEtran.bst files, whereas the Microsoft Word templates are self-contained. Causal Productions has used its best efforts to ensure that the templates have the same appearance.

There is no unique algorithm resulting in high performance, suitable for all kinds of grid computing environment. The major problem being proper assignment of the tasks among the processors. The Genetic Algorithm and LGR method found to improve the performance, advantage being it also considers the communication costs.

The Grid has the characteristics of heterogeneity and dynamicity and these features are distributed hierarchically and locally in many cases. Current Grid resources are usually distributed in a clustered fashion. Resources in the same cluster usually belong to the same organization and are relatively more homogeneous and less dynamic in a given period. The Grid scheduler algorithm may take multiphase or multilevel strategies, a Grid scheduler can first find a coarse scheduling the global Grid and then a fine schedule in a local cluster.

In cases where the knowledge of current load and network conditions is at hand, Ant Colony Optimization algorithm is found to produce good results.

This paper has described an overview of most of the important algorithms so that the reader will be able to find the suitable algorithm in different grid computing environment.